**Response to sudden surge in human movement by an urban-adapted animal.**

Debottam Bhattacharjee[1,2] and Anindita Bhadra[1*]

**Affiliation**:

[1] Department of Biological Sciences, Indian Institute of Science Education and Research Kolkata, Nadia, West Bengal, India.

[2] Animal Ecology Group, Utrecht University, the Netherlands.

[*]**Address for Correspondence**

Behaviour and Ecology Lab, Department of Biological Sciences,

Indian Institute of Science Education and Research Kolkata

Mohanpur Campus, Mohanpur, Nadia

PIN 741246, West Bengal, India.

[*]Corresponding author

*E-mail:* abhadra@iiserkol.ac.in (AB)

*tel.* +91-33 6136 0000 ext 1223

*fax* +91-33-25873020

**Abstract**

Interaction with its immediate environment determines the ecology of an organism. Species present in any habitat, wild or urban, may face extreme pressure due to sudden perturbations. When such disturbances are unpredictable, it becomes more challenging to tackle. Implementation of specific strategies is therefore essential for different species to overcome adverse situations. Numerous biotic and abiotic factors can alter the dynamics of a species. Anthropogenic disturbance is one such factor that has considerable implications and also the potential to impact species living in the proximity of human habitats. We investigated the response of an urban adapted species to a sudden surge in human footfall or overcrowding. Dogs (*Canis lupus familiaris*) living freely in the streets of developing countries experience tremendous anthropogenic pressure. It is known that human movement in an area can predict the behaviour of these dogs by largely influencing their personalities. In the current study, we observed a strong effect of high and sudden human footfall on the abundance and behavioural activity of dogs. A decline in both the abundance of dogs and behavioural activities was seen with the increase in human movement. Further investigation over a restricted temporal scale revealed reinstated behavioural activity but non-restoration of population abundance. This provides important evidence on the extent to which humans influence the behaviour of free-ranging dogs in urban environments.

**Introduction**

Environmental disturbances can impact organisms living in any habitat. Changes in both biotic and abiotic factors may lead to such perturbations affecting the dynamics of a species [1]. Organisms need to overcome adverse situations by building strategies primarily through changing behaviour, physiology, and morphology. Unlike predictable factors, sudden and short-lived unpredictable perturbations are difficult to anticipate and can hamper the survival of a species to a great deal. Organisms, especially animals, need to react to such sudden changes immediately by rapid modifications in behaviour and/or physiology [2]. Some classic examples of environmental perturbation are situations of a flood [2–4] and severe storm [5]. Human-induced disturbances, on the other hand, can be direct, short-lived with potential long-term implications. For example, increased nocturnality in different animal taxa globally is a marked behavioural change due to human disturbances [6]. Animals that share the same environment with humans can help measure the effects of such disturbances. Street dogs found largely in the countries of the global south represent an excellent model system for investigation along these lines. Such studies can be useful in understanding both the effect of human-mediated changes and subsequent responses by the species.

While the role of anthropogenic disturbances in wildlife conservation has received well-deserved attention, data regarding their potential ramifications on urban-adapted animals are scarce. Such studies will be instrumental in providing significant insights into animal population management plans in urban areas, especially in the global south, that are experiencing rapid urbanization. Free-ranging dogs — 'domesticated' dogs living on the streets of the global south — form an integral part of the urban ecosystem [7]. Their activities are not under direct human supervision barring complete victual dependence [8,9]. However, humans are found to be a significant factor influencing the social interaction networks of dogs, especially with regards to initiating behaviours [10]. Humans are responsible for negatively impacting these dogs, often causing mortality [11]. Therefore, free-ranging dogs are subject to human-induced disturbances to a great extent. Consequently, adaptive strategies have been observed in their behavioural responses. For example, they can respond to human social cues in a situation-specific manner [12,13], and show general aversion towards making direct physical contact with unfamiliar humans [14]. However, no attempts hitherto have been directed towards measuring their population-level response to sudden human-induced disturbances, like overcrowding.

Human flux, defined by the movement of humans in an area/minute, was found to predict the varying sociability levels of free-ranging dogs in different 'microhabitats' [15]. For example, dogs in completely residential areas (low flux zones) were shy and/or fearful, possibly due to a lack of human socialization. On the contrary, dogs in intermediate (residential areas with small shops and markets) and high (railway and bus stations, big markets) human flux zones were found to be more sociable than their counterparts in the low flux zones. However, no information is available regarding the consequences of a sudden surge in human footfall in an area, as is evident during festivals, fairs, etc., on dog populations. In this study, we measured human flux during a festive season, which also leads to a surge in food resources and quantified its impact on dog populations from the perspective of abundance and behavioural activity. We hypothesized that a sudden surge in human footfall would impact the dynamics of dog populations. Unfortunately, attempts to form specific, informed predictions were hampered by inadequate existing literature coupled with the exploratory nature of the study.

## Methods

**(i) Study area and sampling**

The study was carried out at 13 locations in the cities of Kolkata (22°57'26" N, 88°36'39" E) and Raiganj (25°61'85" N, 88°12'56" E), West Bengal, India **(Supplementary Information**

**S1)**, in three sessions - 'pre-event', 'event', and 'post-event'. A massive surge in the human movement here characterized the event. We considered the largest festival (Durga Puja/Autumn festival 2019, ~7 days) in West Bengal as our event. During the event, large temporary structures (pandals) are built within cities which attract huge crowds. Human footfall in these areas increases manifold during this time. We selected the 13 sites based on pre-existing information regarding famous crowd-puller Durga Puja sites in the two cities. While Kolkata is the biggest and most crowded city of the state with the most number of pandals, Raiganj is a small city with fewer pandals.

We collected the event data during $14^{th} - 19^{th}$ October 2019. Pre-event data were collected one month before ($8^{th} - 13^{th}$ September), whereas, the post-event data were obtained one month after the event ($10^{th} - 18^{th}$ November). Data collection was done within 1000 – 1200 hours intentionally to avoid extreme crowding in the evenings. We prepared maps of the study locations using Google maps (https://www.google.com/maps) to have information of the streets in detail before carrying out the fieldwork. The mean sampling area size was $1.28 \pm 0.24$ sq km ($s^2 = 0.05$) which was calculated using Google Earth (v 9.2.76.4). We walked along the roads, estimated dog abundance, and measured human movement. Sampling of dogs was carried out using the protocol standardized by Sen Majumdar et al. 2014 [9]. Once a dog was sighted, we recorded it's location, sex (by observing genital structures [16]), and behaviour, using instantaneous scans [17]. We collected the GPS data using *Google my maps* application in a cellular mobile device. Finally, we defined human movement or flux for each sampling area as the number of humans or vehicles that crossed a focal site within the area per minute, as mentioned in Bhattacharjee et al. 2020 [15].

**(ii) Data analysis and statistics –** We primarily focused on dog abundance, but also looked at the behaviour of dogs, in terms of activity. Active behaviours included maintenance, vocalization, interaction, and individual-level behaviours (e.g. foraging, walking, etc.), while inactive behaviours included lazing, sleeping, and sitting [18]. To examine the effect of the event session, we first analyzed the pre-event and event data. We conducted a GLMM (Generalized linear mixed model) analysis with Poisson distribution and "log" link function to investigate and validate the effect of sampling session (pre-event/event) on human flux. Further, we carried out GLMM analyses to understand the impact of sampling session on dog abundance and behaviour (activity). We used a Poisson distribution with "log" link function in the model investigating dog abundance. We considered sex (male/female/unidentified) as a control variable in the model exploring activity (active/inactive); binomial error distribution with "logit" link function was used in this model. Study locations were included as random effects in all the models. We further compared the pre-event and post-event data in order to quantify the long-term impact of human flux, if any. We followed the same analytical procedure as earlier, except for a change in the sampling session variable (pre-event/post-event).

Null vs full model comparison was made for all the models. We used AIC values for selecting the 'best-fitted' model [19]. Generalized linear mixed models (GLMM) were performed using "lme4" package [20] of R (Version 3.6.2) [21], using R Studio (version 1.2.5033). Residual diagnostics and dispersion of the models were checked using DHARMa package of R [22]. The alpha level was kept 0.05 throughout the analyses.

**Results**

A marked increase in human flux was noticed during the event ($125.00 \pm 24.60$) as compared to the pre-event ($49.83 \pm 23.52$) session (GLMM: $z = 19.82$, $p < 0.001$, **Figure 1**), as expected. We obtained a significant effect of the sampling session on dog abundance (GLMM: $z = 4.306$, $p < 0.001$). Dog abundance was found to be higher during pre-event ($15.92 \pm 10.64$) than the event (9.84 ± 8.08) session (**Figure 1**). GLMM analysis revealed that dogs were more active during pre-event (~ 51% of the dogs) than the event (~ 33%) session (GLMM: z = 2.704, p = 0.006); however, no effect of sex was found (Male-female, GLMM: z = -0.525, p = 0.59; female-unidentified, GLMM: z = -0.248, p = 0.804).

Human flux was comparable between pre and post-event (47.95 ± 22.48) sessions (GLMM: z = 0.675, p = 0.50, **Figure 1**). However, the abundance of dogs was found to be lower during post-event (11.84 ± 5.32) as compared to the pre-event session (GLMM: z = 2.797, p = 0.005, **Figure 1**). We did not find any effect of sampling session on the activity of dogs between pre and post-event (~ 42% exhibited active behaviour) sessions (GLMM: z = 1.410, p = 0.16).

**Figure 1**

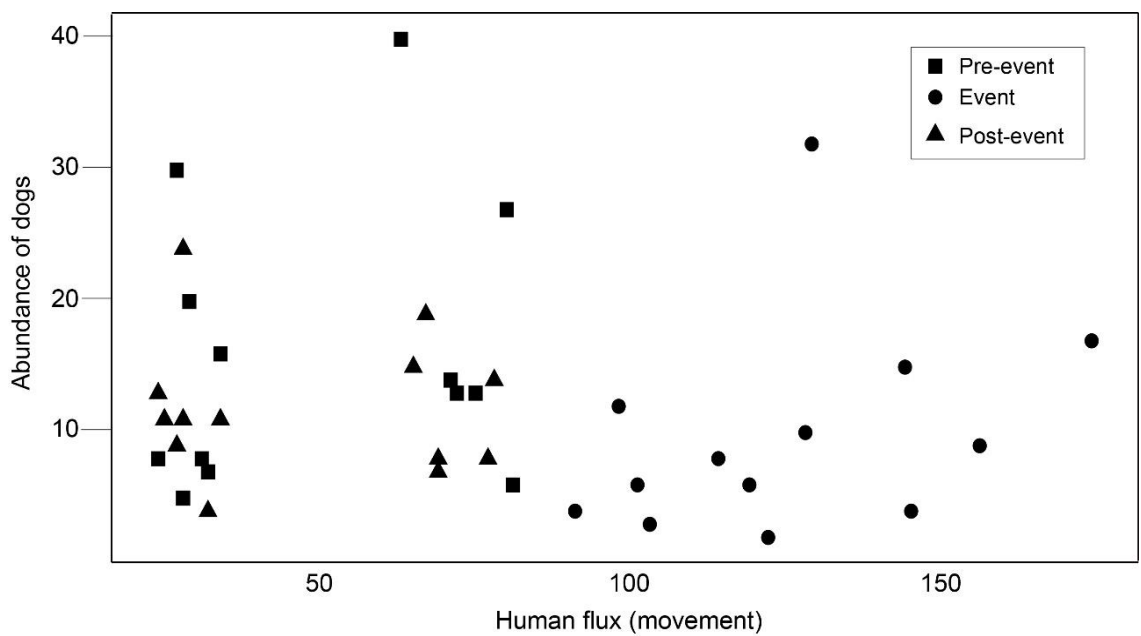

**Discussion**

The results show a strong impact of a sudden surge in human footfall on the dynamics of free-ranging dogs, supporting our hypothesis. A marked decline in the abundance of dogs during the event session, in spite of the increased availability of food, is indicative of increased human disturbances through overcrowding. Also, the reduced behavioural activity of dogs represents the impact of overcrowding on them. Albeit the reinstatement in behavioural activities, non-restoration of population abundance during the post-event session highlight the residual implications of such an event.

Reduced abundance during the event session could be perceived as a temporary migration of dogs from the zone of human disturbance. It appears to be an adaptive strategy for the dogs to displace from such areas to overcome potential adversities. Another probable reason could be the limited access to available food resources. Note that during festive seasons, though the food resources rise manifolds (**Supplementary Information S2**); overcrowding can restrict the utilization of such resources (personal observations). Moreover, overcrowding may hamper the existing access to food resources to the dogs and increase threats from humans while attempting to scavenge. Consequently, to avoid competition for food, dogs might have migrated to nearby areas with less human disturbances. The reduced abundance of dogs in the post-event session, as compared to the pre-event session suggests overcrowding might be a disruptive force impacting the free-ranging dog dynamics significantly.

Reduction in the behavioural activity of dogs during event might stem from restricted access to food and threats from humans. This observation is congruent with a host of laboratory-based studies under controlled conditions that have looked at the effects of feed restriction on animals compared to their *ad libitum* access counterparts. Access to food sources remain severely impeded throughout the seven days of the 'event' owing to the 24X7 overcrowding. Their response to sustained deprivation of access to food could be considered analogous to laboratory animals wherein the latter are known to efficiently reduce energy expenditure by behavioural adjustments (reduced activity) in such conditions [23–27]. Severe disruptions in the immediate environment are also known to instigate stress manifested by physiological changes [28]. As a further extension of this work, levels of cortisol and glucocorticoid measurements could be carried out to quantify effects of such disturbances on dogs. Subsequently, studying dog abundance in the nearby areas with reduced or no human disturbance, and tracking of individual dogs could provide a clearer picture of their migration patterns. Nevertheless, this is the first attempt to look at the effects of direct anthropogenic disturbances on an urban-adapted animal in its natural environment.

**Supplementary Information –**

**S1 – A map of West Bengal (left panel), India showing the 13 study locations (right panel) from the cities of Kolkata and Raiganj (Kolkata –** Maddox Square, Tridhara, Singhi Park, Abasar Pally-Bakulbagan, Mudiali-Shivmandir, Jodhpur Park, Ashar Sangha-66 Pally, Md. Ali Park, College Square, and Bagbazar**; Raiganj –** Chaitali, Merchant Club, and Sudarshanpur.**).**

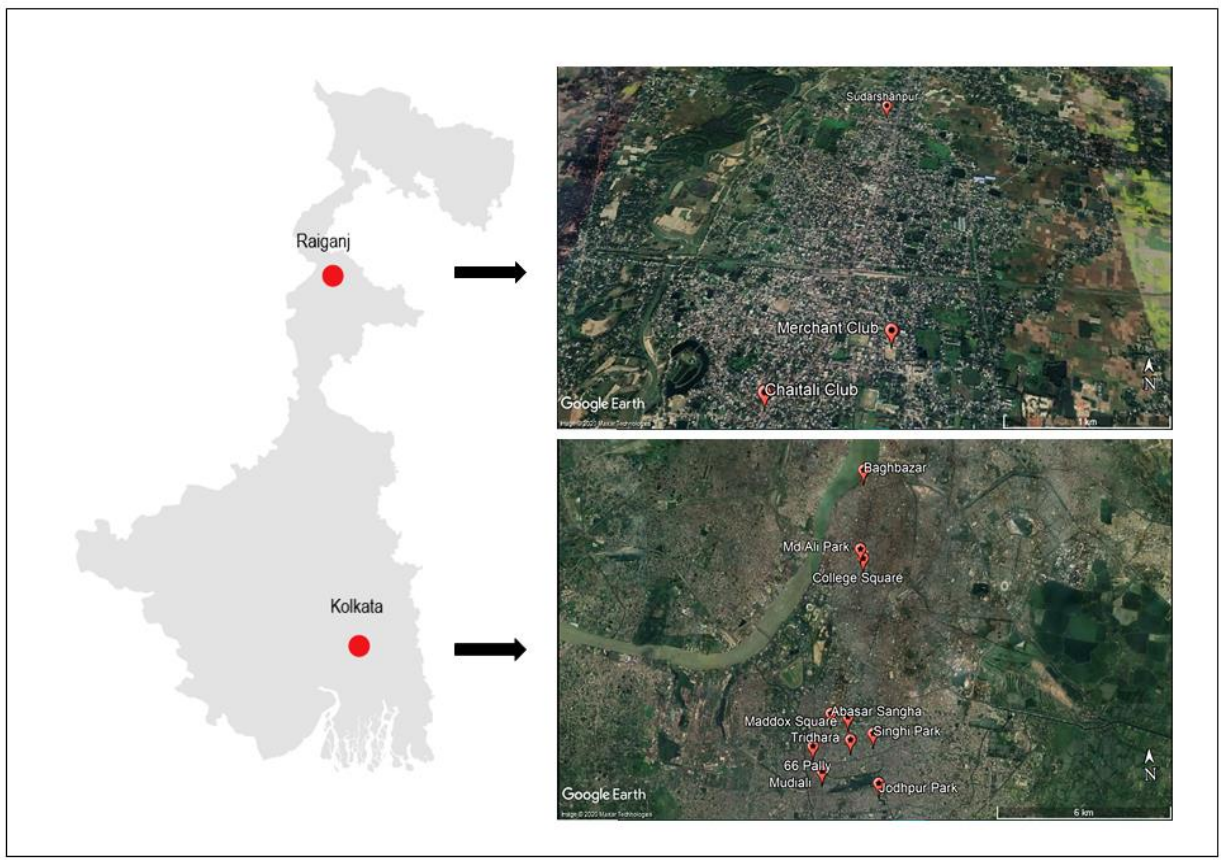

**S2 – Potential food resources available and their scores based on approximate feeding values.** We use these values for quantifying food resources (namely 'Resource Index'). For this study, which did not include longer behavioural observations, we could not distinguish between regular and irregular sources. Resource Index of an area was calculated using summations of all available food resources.

| Type of source | Food source | Description | Score |
|---|---|---|---|
| Regular sources | Meat shop | Open ones only, from where dogs get access to the bits and pieces of meats | 8 |
| | Eatery / Restaurant | Roadside open restaurants/eateries. Restaurants like Subway and McDonald's were not considered as free-ranging dogs do not get access to the leftover foods from such eating places. | 7 |
| | Direct feeding by human | Provision of food directly by humans (may or may not be by means of begging). Restricted to food provisioning from households. | 6 |
| | Open garbage dump | ≥ 2 sq m in size, consisting of wet and dry garbage, including leftover food by humans. It can accommodate several individuals. | 5 |
| | Tea shop | Open ones only. Tea shops may also offer bread/biscuits. | 4 |
| Irregular sources | Direct feeding by human | Occasional provisioning of food by humans only by means of begging. | 3 |
| | Grocery / sweet shop/ bakery | All these shops come under one umbrella as they offer direct food provisioning to dogs occasionally. | 2 |
| | Household or eatery dustbins (without cover). | ≤ 0.5 sq m in size consisting of wet and dry garbage. It can accommodate up to two individuals at any given time. | 1 |

Based on the values of Resource Index, we found a significant increase in the resources during the session of event (Wilcoxon paired-sample test: T = 3, N = 13, p = 0.001; see figure below) as compared to pre-event. We found no difference between the pre-event and post-event session Resource Index values (Wilcoxon paired-sample test: T = 35.500, N = 13, p = 0.76; see figure below).

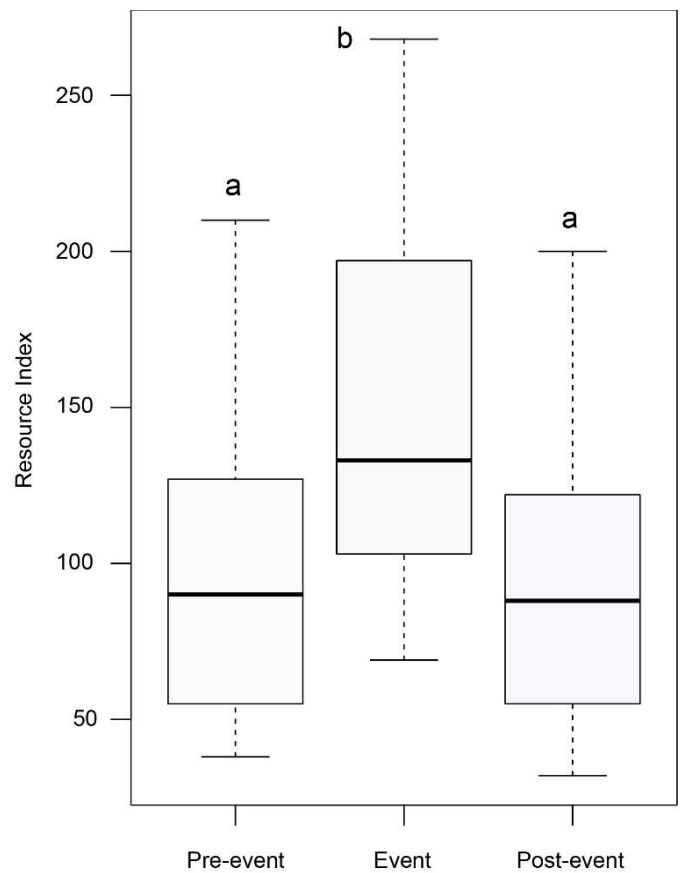